\documentclass[11pt, a4paper, twoside]{article}
\usepackage[utf8]{inputenc}
\usepackage{amssymb}
\usepackage{color}
\usepackage{youngtab}
\usepackage{graphicx, fancybox,amsfonts, amsmath, amsthm, relsize, simplewick, float,subcaption,mathtools,cite}
\usepackage[colorlinks=false,linkcolor=green]{hyperref}
\usepackage{relsize}
\usepackage{algorithm,algpseudocode}
\def\bra #1 {{\left\langle #1 \right|} }
\def\ket #1 {{\left| #1\right\rangle} }

\author{\textbf{Hasina Tahiridimbisoa Nirina Maurice, Yabebal Tadesse}}
\usepackage[left=3.5cm,right=3.5cm,top=3.5cm,bottom=3cm]{geometry}

\newcommand{\be}{{\mathbf{e}}}

\newcommand{\Bmax}{B_{max}}
\newcommand{\bq}{\mathbf{q}}

\DeclareMathOperator{\Tr}{Tr}
\DeclareMathOperator{\T}{T}
\newcommand{\xo}{x_1}
\newcommand{\xto}{x_2}
\newcommand{\xtr}{x_3}
\newcommand{\xfo}{x_4}
\begin{document}
%%%%%%%%%%%%%%%%%%%%%%%%%%%%%%%%%%%%%%%%%%%%%
\pagenumbering{gobble}

\Large
 \begin{center}
\textbf{{Symmetric group and the Axelrod model for dissemination of cultures}} \\

\hspace{10pt}

% Author names and affiliations
\large
Nirina M. Hasina Tahiridimbisoa${}^{a,}$\footnote{nirina@aims.ac.za}, Yabebal Tadesse${}^{b,}$\footnote{yabi@aims.ac.za}\\

\hspace{10pt}

\small  
${}^{a,b}$African Institute for Mathematical Sciences, \\

${}^{a,b}$Department of Applied Mathematics Stellenbosch University \\

\end{center}

\hspace{10pt}

\normalsize
\hrule
\vspace{1cm}
\textbf{Abstract:}
We consider the model proposed by Axelrod for dissemination of cultures on a 2-dimensional squared lattice. We review this model from an analytic point of view. We define $\left\langle s(t)\right\rangle$ to quantify possible culture configurations at time $t$ in a society. Typical initial culture configurations of this model are characterised. Equation of motion in terms of $\left\langle s(t)\right\rangle$ is derived. We study the graph of development of this Axelrod system toward to its culture configurations equilibrium. Generically, we observe that this model undergoes three phases of development. We give a quantitative explanation about these three different phases of development.
 Keeping up with this Axelrod model, we characterize its culture configurations space at equilibrium point where $\left\langle s(t_{\text{eq}})\right\rangle = 1$. This space is called monoculture space. Understanding this space is equivalent to restrict to the space of culture configurations from one individual in the model. This individual culture space is identified to the space $V_N^{\otimes F}$ up to isomorphisms. Action of the permutation group $S_N$ on the space $V_{N}^{\otimes F}$ is considered. Under this action, the observable $\left\langle s(t)\right\rangle$ is an invariant of the Axelrod system. We explore this symmetry and classify the different inequivalent classes of culture configurations composing the monoculture space. To achieve this, we consider the case $N\geq F$. We propose techniques from group representation theory to perform this classification. The inequivalent classes of culture configurations are indexed by the Dickau diagrams which are associated to the Bell number $B_F$. A concrete example with $F=4$ and $N\geq 4$ is considered for a full illustration of our analysis.
  \vspace{.5cm}
 \hrule
 \vspace{.3cm}
\newpage
\pagenumbering{arabic}
%\maketitle
%\section{Nice properties about the Axelrod Model for cultural dissemination:}
%Nice things about the proposed model for dissemination of cultures by Axelrod are
%\begin{enumerate}
%\item\textbf{Abstraction of Culture:} The model select finite set of independent features or attributes from an \emph{infinite set} of possible features that people can influence between each other. This was defined as an abstraction of culture in the model. In addition, each attributes take values in a finite set which can be mapped one-to-one to a finite set of integers. In this way, one can study the convergence of social influence.
%\end{enumerate}
%\section{New things to be consider:}
%\begin{enumerate}
%\item \textbf{What if we allow infinite distance range of interaction?}
%\item \textbf{It is important that this model is irreversible, unlike what might happen in reality}
%\item \textbf{What if we allow feature dependence and there is self interaction}
%\end{enumerate}
\section{Introduction and motivation}
Constructing a mathematical model of human society or any other biological society turns out to be very challenging. In fact, their complexities and dynamical properties appear to be far more difficult than studying the dynamics of the elementary particles that make up the building blocks of our universe. One explanation of this is the lack of evidence for a fundamental theory\footnote{In elementary particle physics, it is nowadays understood that the correct framework is summarized into Quantum Field Theories or just QFTs} towards to this study. The best approach so far is to try different ideas from wide range of frameworks and solve specific questions of interest. Examples of these are found in \cite{homans61,triandis94,gilbert95,epstein96,Axelrod97,Jin01,miller01,newman02,grimm05,goldstone05,jasmine06,vasques07,goldstone08,davidletal09,buchanan} and their references.

 However, among the existing literatures on social science it turns out that a common background idea corresponds to certain idea encountered in the statistical physics framework. We call this as a social science/statistical physics correspondence. This idea was already exploited long before in \cite{steven99,buchanan}. A recent survey trying to establish this connection among a wide list of results is found in \cite{CCFVL09}. In support of this correspondence, inspired by \cite{Axelrod97}, the authors of \cite{gandica13,genzor15} build thermodynamic models of social influence. 
Following these authors, we also support that there is indeed such a correspondence. We take it seriously and exploit to study a ``culture" and its dynamical spread motion in a society.  
We adopt the idea that a ``culture" in human society is an emergent property of a population. This definition follows from our statistical physics point of view such that a culture must be an average statistic of individual behaviours in the population. For a relevant and introductory background in statistical physics we refer to the David Tong lecture notes \cite{DavidTong12}.

In this work, a question we ask is how to construct a model of human society that abstracts their culture and describe it mathematically? To answer this question, a priori one has to know first what are the possible states of individual behaviours? In fact, answering this last question makes this problem very challenging. That is because even the simplest existing model \cite{Axelrod97} we are interested in, leads to a tremendous number of culture configurations in the system. In fact, even for ``small" system - see equation \eqref{confOmega} - the expected number of culture configurations is far more larger than the estimated numbers of atoms in our universe. This is a serious challenge where one is intimidated in tackling this problem analytically. However, this is where we use relevant techniques from statistical mechanics. That is because such a challenge is always encountered in this framework. In addition to this, we also explore the power of having the group $S_N$ as a symmetry of this Axelrod model. 

To our knowledge, the novel contributions of our work are as follows:
\begin{itemize}
	\item Exploring the idea of the social science/statistical physics correspondence in reviewing analytically Axelrod model \cite{Axelrod97} but keeping its original principle.
	\item The use of the $S_N$ symmetry present in the system to analyse the culture configurations space in this model.
	\item We propose a simple algortihm to decompose $V_N^{\otimes F}$, which employs Schur-Weyl duality between $S_N$ and $S_F$. This is related to partition algebras as described in \cite{vfrjones93,halverson05, zajj14}. This decomposition of $V_N^{\otimes F}$ gives the inequivalent classes of culture configurations composing the monoculture space.
\end{itemize}
An analytic review of the Axelrod model specific to the one-dimensional model is found in \cite{lanchier12}. However, our analytic review here is to take the model from its first principle. 

Apart from our interests in this work, because of the techniques we propose, it is worth to point out the following. There is a connection of part of our results with other totally different field in physics.  Schur-Weyl duality, the space $V_N^{\otimes F}$ and the symmetric group are three major abstract concepts that appear in our study and commonly encountered in the study of the AdS/CFT correspondence \cite{maldacena97}. A first suggestion exploiting these abstract concepts and using representation theory to study this duality was introduced in \cite{corley01}. For necessary backgrounds on this topic, we refer to the article \cite{dmk11}. Many results have been published in supporting this idea and here is a few list of the recent ones including some of their references, \cite{dMKH16,NMHT,dmk17,dmk18}. Accordingly, there is a non-trivial overlap with certain techniques used in these literatures with what we employed here. So we refer readers to these materials for relevant backgrounds needed to achieve the relevant parts of our results. Moreover, we believe the decomposition we report here maybe useful for future studies in this direction. In fact, this is already the case in the recent published article \cite{dmk17}. 
 
Extra comment motivating why we think the social science/statistical physics correspondence should be powerful. First, the successes in the statistical physics framework are based on the fact that physical observables are mathematically well defined. Furthermore, they enjoy many symmetries so that finding their spectrum can be turned into well defined mathematical statements. Under this correspondence, our insight is to treat at the same footing individuals composing a society as elementary particles that compose certain physical system\footnote{This statement ignore the quantum behaviours of particles. We only consider them at the classical level.}. In this way, we try to apply these successful key ideas from the statistical physics side to the context of studying cultures in the society. In this pursuit, we give a quantitative aspect of a culture and its dynamic. 

We propose the quantity $\left\langle s(t)\right\rangle$ measuring the average similarity of pairs of individuals in the population. A simple physical interpretation of the observable $\left\langle s(t)\right\rangle$ is that it can be used to measure at time $t$ the departure of a culture configuration in a society to be a monoculture. A culture configuration in a society is monoculture if all individuals composing this society are sitting in the same individual culture state. If this is not the case, the culture configuration in this society is called a multicuture state. Our definitions of the previous technical terms are in agreement with the following literatures \cite{castellano00,rodriguez10,gandica13,genzor15}. 

To achieve our goal, as stated earlier, we consider the model \cite{Axelrod97}, proposed by Axelrod for dissemination of culture in human society without any external authorities. This model uses the idea of homophily which turns out to be fundamental to individuals in a society. Homophily is a phenomenon between individuals to only form a bond or associate with each others if they share a non-trivial similarity and tend to increase it. This model is classified as an agent-based modeling. The methodology adopted in the original paper relies on simulating the interaction of individual agents and then observes the emergence of global properties of the system. Data were collected from many simulations to perform the analysis of the model itself.
However our approach here is rather based on an analytic analysis. This is one of the main motivation in this article. Having an analytic understanding of the model helps to generalize the model when the system considered is no longer computationally feasible. Despite its simplicity, this model is actually one of the most accepted quantitative model how culture spreads over time. 

This article is organized as follow. In section \ref{Defnot} we give relevant definitions and setup notations to describe the system. We characterize the likelihood of typical culture configurations of the Axelrod system prior to any interaction within itself. We derive the probability distribution of similarities over individual states. Following this, we describe the culture development of the system over time.  In section \ref{SecCCSN} we analyse the space of cultural configurations in the system. This is achieved by understanding individual's culture space and identify it isomorphically to the space $V_N^{\otimes F}$. The action of the group $S_N$ is defined in this section. Some elementary combinatorics fact related to this model are also discussed in this part. Finally, in section \ref{concl} we summarize our findings and describe possible outlooks that might be considered as extension of our results. 

\section{Definitions and notations}\label{Defnot}
Following \cite{Axelrod97} and like many other papers inspired by it, we adopt the proposed definition of the word \emph{culture} to abstract a real culture in a society. The model is constructed by defining individual's culture from a finite set of different and independent attributes $q_i$, where $i$ is a positive integer ranging from $1$ to $F$. Denote this set by $\mathcal{Q}_F$.  Each attribute $q_i\in \mathcal{Q}_F$ is chosen according to the principle that the value it takes can influence others. Example of these attributes are language, religion, political party and other things. 

In what follows, we stress the importance that the $q_i'$s are independent. First this independence means that values that two different $q_i$'s take can not affect each other directly. An illustration where two different $q_i$'s are not necessarily independent is if we consider religion and food as attributes. A priori these two attributes are different. However, it is easy to check in a real society that if someone belongs to a certain type of religion, this attribute usually imposes constraints directly to the type of food that same person may only eat. According to our convention the attributes religion and food are not independent. Hence, these two attributes are not compatible in this model and should not be considered. We will use later this independence of attributes to derive the probability distribution of local degree similarities over individual states. At this stage, it is important to mention that even if this independence of attributes was not mentioned in the original paper the author already takes it into account. Furthermore, we also need to stress the fact that independence of attributes has no contradiction to the following hypothesis. In \cite{Axelrod97}, it is assumed that the effect of a feature $q_i$ on the system depends on the absence or presence of other features $q_j$. In fact, this is where the implementation of the homophily principle that governs the dynamics of the model is about. So we also support this hypothesis but it should not be confounded with the independence of attributes we mentioned earlier. 

 For simplicity let each attributes $q_i\in \mathcal{Q}_F$ take value in a set of $N$ different traits. As a concrete example, let $q_1$ be the attribute associated to a type of religion. The $N$ different traits can be selected from being $\{ \text{Christian}, \text{Muslim}, \text{Buddhist},\cdots\}$.
 Following Axelrod, one can define a bijection, mapping these $N$ traits to the set of positive integers ranging from $1$ to $N$. The power of using these integers will be reflected later in our analysis of the culture configurations of this system. However, a comment is that Axelrod used these integers to be able to facilitate the simulations. But in our work we demonstrate without loosing any physical interpretations that these integers help to define the action of the symmetric group $S_N$ on the system. This action leads to the realization that the group $S_N$ is a symmetry of the quantity $\left\langle s(t)\right\rangle$. Furthermore, the classification of the space of culture configurations in the population is simplified by these integers. 
  
 To summarise the above descriptions on the individual's culture, we now introduce a culture state of an individual $I$ 
 \begin{equation}\label{microstate}
{\ket{\bq} }_I \equiv {\ket{q_1q_2\cdots q_{F-1}q_F} }_I.
\end{equation}
Since each $q_i$'s can take values in $\{1,2,\cdots,N\}$, it follows that the number of microstates accessible by one individual is
\begin{equation}\label{eq:nmicrostates}
	\Omega_I = N^F.
\end{equation} 
The Dirac \emph{ket} notation of the state is just a notation, and so it has nothing to do with quantum mechanics. The reason we only use it here is just to have a better representation of an individual state. 
Following \cite{Axelrod97}, to make up the system, the model considers a geographical distribution of agents on a two-dimensional square lattice of size $L\times L$.
See table \ref{table1} below for a typical initialization of a system configuration.
\begin{table}[H]
	\centering
	\caption{Illustration of a typical initial configuration with $F=5$, $N=10$ and $L=10$. It uses the integers $0,1,\cdots ,9$ instead of $1$ to $10$.}\label{table1}
	{\scriptsize
		\text{\textbar}46317\text{\textbar}57215\text{\textbar}37500\text{\textbar}80227\text{\textbar}13364\text{\textbar}53540\text{\textbar}84835\text{\textbar}92036\text{\textbar}16595\text{\textbar}34302\text{\textbar} ~\\
		\text{\textbar}02305\text{\textbar}25283\text{\textbar}07264\text{\textbar}76387\text{\textbar}13680\text{\textbar}05932\text{\textbar}49216\text{\textbar}59984\text{\textbar}85216\text{\textbar}67325\text{\textbar} ~\\
		\text{\textbar}61998\text{\textbar}59470\text{\textbar}63884\text{\textbar}60829\text{\textbar}16146\text{\textbar}63117\text{\textbar}36062\text{\textbar}02974\text{\textbar}00047\text{\textbar}07716\text{\textbar} ~\\
		\text{\textbar}76076\text{\textbar}17738\text{\textbar}26408\text{\textbar}27114\text{\textbar}16679\text{\textbar}48805\text{\textbar}63941\text{\textbar}85828\text{\textbar}05781\text{\textbar}86808\text{\textbar} ~\\
		\text{\textbar}71998\text{\textbar}09225\text{\textbar}20536\text{\textbar}53472\text{\textbar}71024\text{\textbar}66115\text{\textbar}26271\text{\textbar}82997\text{\textbar}34706\text{\textbar}00832\text{\textbar} ~\\
		\text{\textbar}63917\text{\textbar}40374\text{\textbar}59187\text{\textbar}11198\text{\textbar}09243\text{\textbar}86905\text{\textbar}95275\text{\textbar}65085\text{\textbar}50814\text{\textbar}83458\text{\textbar} ~\\
		\text{\textbar}01242\text{\textbar}81728\text{\textbar}75428\text{\textbar}14405\text{\textbar}69990\text{\textbar}19809\text{\textbar}20541\text{\textbar}33572\text{\textbar}34125\text{\textbar}80097\text{\textbar} ~\\
		\text{\textbar}24375\text{\textbar}15114\text{\textbar}30587\text{\textbar}29830\text{\textbar}57592\text{\textbar}17560\text{\textbar}97670\text{\textbar}15430\text{\textbar}71994\text{\textbar}81084\text{\textbar} ~\\
		\text{\textbar}85585\text{\textbar}03874\text{\textbar}30655\text{\textbar}88627\text{\textbar}11036\text{\textbar}26252\text{\textbar}85558\text{\textbar}52786\text{\textbar}90468\text{\textbar}16155\text{\textbar} ~\\
		\text{\textbar}39126\text{\textbar}48158\text{\textbar}53209\text{\textbar}90686\text{\textbar}69573\text{\textbar}57734\text{\textbar}82503\text{\textbar}29802\text{\textbar}22136\text{\textbar}44851\text{\textbar}. \\ }
\end{table}
\noindent
At this point, it is useful to give a quick summary of the main parameters in the model
\begin{itemize}
	\item[-] $F$ counting the number of features or attributes that one can influence others. 
	\item[-] $N$ counting the number of traits that each attribute can take. 
	\item[-] $L$ measuring the size of population on a two-dimensional square lattice.
\end{itemize}
Given these parameters, the total number of cultural configurations of the system is 
\begin{equation}\label{confOmega}
	\Omega = N^{F\times L^2 }.
\end{equation}
To see how enormous this number is, we consider the case as in table \ref{table1}, $N=10, F = 5$ and $L=10$. For this system the number of culture configurations is $\Omega = 10^{500}$. 

Proceed to the definition of the local degree of similarity between two individuals in the system. Toward to this, we like to think of the attributes $q_i$'s as vectors in a $N-$dimensional vector space. Let $V_N$ be this vector space. The canonical basis vectors of $V_N$ is \[\mathcal{B}_N \equiv \{\be_1,\be_2,\cdots,\be_N\},\] where the $\be_j$'s are column vectors with their $k^\text{th}$ components 
\begin{equation}\label{eq:BN}
(\be_j)_k = \delta_{jk}.
\end{equation} 
Now, we define
\begin{equation}
	q_i \mathlarger{\rightleftharpoons} \be_{q_i}\quad \Rightarrow \quad \ket{\bq} = \ket{\be_{q_1}\be_{q_2}\cdots \be_{q_F}} .
\end{equation}
It follows that an individual state $\ket{\bq} $ can be identified as an element of the space $V_N^{\otimes F}$. This space is the tensor product of $F$ copies of $V_N$. However, we will save this discussion later in section \ref{SecCCSN}. Return to our derivation of the local degree similarity. Note that for each state ${\ket{\bq} } $ there is a unique matrix $E_{\bq} $ associated to it. The dimension of this matrix is $N\times F$ and its entries are
\begin{equation}\label{matrixstate}
	\left[E_{\bq}\right]_{ab} = \delta_{aq_b}, \quad 1\leq a\leq N \quad\text{and}\quad 1\leq b\leq F.
\end{equation}
Concretely, consider $N=3$, $F=5$, a state $\ket{\bq} = \ket{\be_1\be_1\be_2\be_3\be_2} $ is uniquely associated to
\begin{align}
E_\bq &= \mathsmaller{\begin{bmatrix} 1 & 1 & 0 & 0 & 0 \\ 0 & 0 & 1 & 0 & 1 \\  0 & 0 & 0 & 1 & 0 \end{bmatrix}}.
\end{align}
From the above definitions, the local degree of similarity between two individuals $I$ and $J$ with their respective individual states ${\ket{\bq} }_{I}$ and ${\ket{\bq} }_J $ is\footnote{Here $E^{\T}_{\bq}$ is the matrix transpose of $E_\bq$. $\Tr(\cdot)$ is just taking the trace of a square matrix.}
\begin{equation}\label{similarityI}
	s_{IJ} = \frac{1}{F}\Tr\big(E^{\T}_{\bq_I} E_{\bq_{J}}\big).
\end{equation}
\subsection{Typical initial configuration}
Determining the typical initial configuration of the system is important in order to understand the dynamic development of the system.
An initialization of the Axelrod system is to generate individual states from a uniform random process. This follows from a natural assumption that prior to any interaction\footnote{We will describe this in the section studying the dynamics of the system.} within the system, individuals are equally likely to be in any microstates defined in \eqref{eq:nmicrostates}. To characterize a generic initial configuration of the system we derive the probability distribution of similarities\footnote{To be precise we refer to the local degree of similarities between pairs of individuals in the system. From now on we always use this as a short-cut notation.}. To achieve this, consider two different agents $I$ and $J$ having respectively the states ${\ket{\bq} }_I$ and ${\ket{\bq} }_J$. Use equation \eqref{similarityI} to define $n_{IJ}=\Tr\big(E^{\T}_{\bq_I} E_{\bq_{J}}\big)$, counting the number of identical shared traits between these two states. Let $P(s_{IJ})$ be the probability that agents $I$ and $J$ have the fraction of similarity $s_{IJ} = \frac{n_{IJ}}{F}$. We find
\begin{equation}
P(s_{IJ}=\mathsmaller{\frac{n_{IJ}}{F}}) =  \binom{F}{n_{IJ}}\left(\frac{1}{N}\right)^{n_{IJ}}\left(1-\frac{1}{N}\right)^{F-n_{IJ}}.
\label{eqproba}
\end{equation}
In the analytic review of the one-dimensional model \cite{lanchier12}, a different approach to derive this probability distribution is proposed.
A comment about equation \eqref{eqproba} is that it is independent to the lattice size parameter, thus it is  independent to the size of the population. In this way, it is clear that equation \eqref{eqproba} applies to any system other than the squared lattice system. This is expected since by definition we define it locally between two randomly selected individuals.

To prove \eqref{eqproba}, focus on a single attribute $q_k$ between the two agents. A straightforward counting implies that the probability such that $q_k\big|_I = q_k\big|_J$ is exactly $\frac{1}{N}$. Continue with this single attribute $q_k$ and identify $q_k\big|_I = q_k\big|_J$ as a success or otherwise a failure with their respective probabilities $\frac{1}{N}$ and $(1-\frac{1}{N})$. Now, given the two states ${\ket{q_1q_2\cdots q_F} }_I$ and ${\ket{q_1q_2\cdots q_F} }_J$. In terms of this success/failure picture, $n_{IJ}$ is counting the number of successes from pairwise identification of the $q_k\big|_{k=1,\cdots,F}$'s in these two states. These demonstrate that the probability of similarity between two randomly selected individuals $I$ and $J$ must be the binomial distribution given in equation \eqref{eqproba}.

\begin{figure}[H]
	\centering
	\includegraphics[scale=.65]{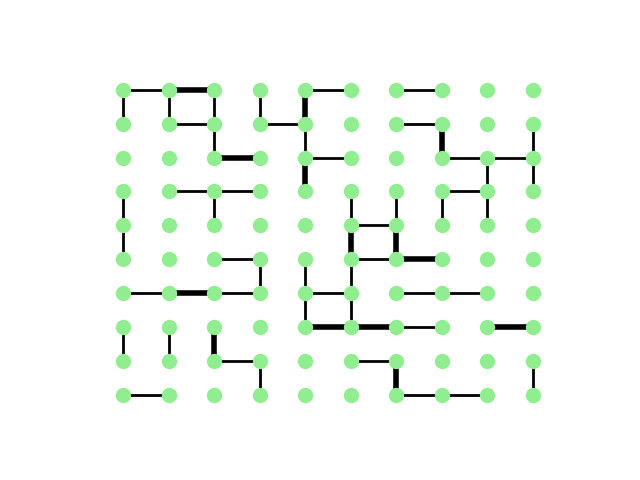} 
	\caption{Gridgraph visualization of the similarities between adjacent sites. In this gridgraph the thicker the edge the bigger the local similarities $s_{IJ}$ is close to $1$. No edge signifies similarity is equal to zero.}\label{typicalGraph}
\end{figure}

In stead of using table \ref{table1}, it is very useful to have a better pictorial visualization of the system. Inspired by \cite{lanchier12}, we propose to use weighted and undirected gridgraph $G_{L,L}$, as illustrated in the figure \ref{typicalGraph}. In fact, figure \ref{typicalGraph} is exactly the gridgraph associated to the system given at table \ref{table1}. The nodes in the graph represent the individuals in the system. Accordingly, each nodes can be in any of the microstates defined in \eqref{microstate}. The vertices of $G_{L,L}$ are labelled by the coordinates of the lattice. The weights of the edges are exactly equal to the the similarity $s_{IJ}$ defined in equation \eqref{similarityI}.

Above, we note that we only take into account similarities between immediate nearest-neighbours. These nearest-neighbours do not take into account the diagonal sites but only those illustrated in figure \ref{typicalGraph}. In this work, our study is only focused on these immediate nearest-neighbours and will leave any possible long range interaction for a future project. 
Motivated by the gridgraph picture of the system, it is natural to use the lattice-coordinates to index each individuals. In this way, the degree similarity in equation \eqref{similarityI} becomes\footnote{For nearest-neighbours only, the restriction on $k$ and $l$ in terms of $i$ and $j$ is $|k+l-i-j|=1$.}
\begin{equation}\label{simijkl}
s_{ij;kl} = \frac{1	}{F}\Tr\big(E^{\T}_{\bq_{ij}}E_{\bq_{kl}}\big),
\end{equation}
where respectively $i$ and $j$ (or $k$ and $l$) locate the row and column positions of an individual on the square lattice. $(i,j) = (1,1)$ corresponds to the first vertex at the top-left in figure \ref{typicalGraph}.
\subsection{Dynamics of the system}
Based on numbers of existing papers inspired by the homophily mechanism implemented by Axelrod in his model, the number of steps in describing their algorithm varies from one to another. However, since as part of our goal we rather give a bit of details on the algorithm governing the dynamics of the model.
The system evolves according to the following steps:
\begin{algorithm}[H]
	\caption{: Detailed algorithm governing the dynamics of the system}\label{Alg1}
	\begin{algorithmic}[1]
		\Statex \textbullet~\textbf{Parameters:} $N, F, L,  \ket{\bq} , t \in \mathbb{N}$ and $s_{IJ}(t)$.
		\State Step-1: Set $t=0$ and initialize individual states in the system to give a configuration on the lattice.
		\State Step-2: Pick at random uniform an individual $I$ sitting in its state ${\ket{\bq}(t) }_I $.
		\State Step-3: Choose at random uniform one of its nearest-neighbours sitting at its state ${\ket{\bq}(t) }_J $, and evaluate $s_{IJ}(t)$.
		\State Step-4: Check the following 
		\begin{itemize}
			\item[i-] if $s_{IJ}(t) \neq 0$, then update the state so that ${\ket{\bq(t)} }_I\rightarrow {\ket{\bq(t+1)} }_I $ and return to step 2. 
			\item[ii-] if $s_{IJ}(t) = 0$, return to step 2.
		\end{itemize}
		\State Step-5: Stop when all pairs of nearest-neighbours - $\{IJ\}$ as illustrated in figure \ref{typicalGraph} - have either $s_{IJ} = 0$ or $s_{IJ} = 1$.
	\end{algorithmic}
\end{algorithm}
\noindent
At step-4, the original model \cite{Axelrod97} actually proposed that even if $0<s_{IJ}<1$, the update of the state ${\ket{\bq(t)} }_I$ may not happen\footnote{This is what one would possibly expected in a real society.}. However, we found this is just causing a time delay for the system to reach its equilibrium configuration. For simplicity we ignore this time delay and step-4 i- always happens as long as $0<s_{IJ}(t)<1$. The update of state $\ket{\bq(t)}  \rightarrow \ket{\bq(t+1)} $ is now explained. Given the two states ${\ket{q_1q_2\cdots q_F} }_I$, ${\ket{q_1q_2\cdots q_F} }_J$ at time $t$ where interaction has to happen, the agent $I$ adapts one of its attribute by equating it to the value of the same attribute in $J$. This attribute is chosen at random uniform from the list of attributes that has no overlap\footnote{Another way to say this is to consider the list of attributes where there is disagreement between the two states.} with those in the state ${\ket{\bq(t)} }_J$. 
A simple explanation to  all the random uniform choices in the above discussion is explained as follow. We treat things equally likely so that any preferential attachment to one item in a list of possible values is not present. This is the most natural and democratic choice we can make since we ignore any information about individuals and their respective states in the system. 

Next, we only consider system that has boundary so that the agents at the four corners of the lattice have two nearest-neighbours. The remaining agents at the boundary have three nearest-neighbours. Otherwise the interior agents have four. It follows that the maximal number of bounds for a square lattice of size $L$ is $\Bmax= 2L(L-1)$. Now, we consider the quantity
\begin{equation}
\label{eqsbar}
\left\langle s(t)\right\rangle  = \frac{1}{\Bmax}\sum_{i,j = 1}^{L-1,L}\sum_{k=0}^{1} s_{ij;(i+k),(j+1-k)}(t),
\end{equation}
which defines the average similarity of the nearest-neighbours pairs in the system at time $t$. One uses it as a metric to quantify how likely a culture in the population is close to be monoculture, i.e. $\left\langle s(t)\right\rangle = 1$. Recall that monoculture system signifies that all agents share exactly the same state.
Note that at a fixed time $t$, a value of $\left\langle s(t)\right\rangle$ does not give a unique culture configuration but instead a set of different culture configurations. This can be argued using the gridgraph illustrated in figure \ref{typicalGraph}. This gridgraph representation of the system discards the actual states of the vertices (which are the individuals). Only it shows the thickness of the edges which code the value of $s_{IJ}(t)$ at each time $t$. 

To understand the set of culture configurations at time $t$ for a given $\left\langle s(t)\right\rangle$, we propose to use the action of the permutation group $S_N$ defined in section \ref{SecCCSN}. Even if a single value of $\left\langle s(t)\right\rangle$ is not unique to a configuration, we can still study the generic evolution of $\left\langle s(t)\right\rangle$ by plotting it versus time. In this way, the graph of $\left\langle s(t)\right\rangle$ can be used to study the motion of the system from an initial configuration until it reaches its equilibrium. Equilibrium configuration is reached if there is no possible interaction to happen within the system. 

Now, we focus on the analytic derivation of $\left\langle s(t)\right\rangle_{t=0}$ for a generic initial culture configuration. Use the probability distribution \eqref{eqproba} to find
\begin{equation}\label{sbart0}
\left \langle s(t)\right\rangle_{t=0} = \frac{1}{N}.
\end{equation}
It appears that $\left \langle s(t)\right\rangle_{t=0}$ is independent of the number of attributes $F$ and the lattice size $L$ related to the size of the population in the system. However there is a fluctuation around this generic value, due to the randomness implemented in the initialization of the system. Our theoretical study demonstrates that $\left \langle s(t)\right\rangle_{t=0}$ depends only on $F$ and $L$ through this fluctuation. To derive these, we refer to Appendix \ref{AppendixA} and use the ergodic hypothesis and the laws of large number.\footnote{These are crucial ideas and often relevant for the derivation of macroscopic quantity in statistical mechanics.} We find that $\left \langle s(t)\right\rangle_{t=0}$ follows a Gaussian distribution that peaks at $\frac{1}{N}$ and a standard deviation $\Sigma_{\left \langle s(t)\right\rangle_{t=0}}\sim\frac{1}{\sqrt{\Bmax}}$. 

To complete our study of the dynamics of the system, we now derive the equation of motion for $\left\langle s(t)\right\rangle$. To achieve this, focus on the interior sites of the lattice. When an interaction happens between time $t$ and $t+1$, one of the following 7 cases of scenario must happen. In terms of $\left\langle s(t)\right\rangle$ we have 
\begin{equation}\label{eq:motion}
	\left\langle s(t+1)\right\rangle = \left\langle s(t)\right\rangle + \delta_{t}, \quad\delta_t\in \mathsmaller{\left\lbrace -\frac{2}{F},-\frac{1}{F},0,\frac{1}{F},\frac{2}{F},\frac{3}{F},\frac{4}{F}\right\rbrace}.
\end{equation} 
It is clear in this equation that $\delta_t$ expresses the change in $\left\langle s(t)\right\rangle$ at time $t$.
It turns out that the appearance of the two scenario with the negative signs is counter intuitive and against the homophily principle in this model. However, we now argue that they must happen in the model when we look at the system globally\footnote{Note that the study of $\left\langle s(t)\right\rangle$ for a system is already taking into account its global property.}. To achieve this, it is enough to consider the two extreme cases and illustrate them in terms of examples.

 Consider the first case where $\delta_t=-\frac{2}{F}$. At the immediate nearest-neighbours, the following picture is a possible explanation of this 
\begin{figure}[H]
	\centering
	\includegraphics[scale=.7]{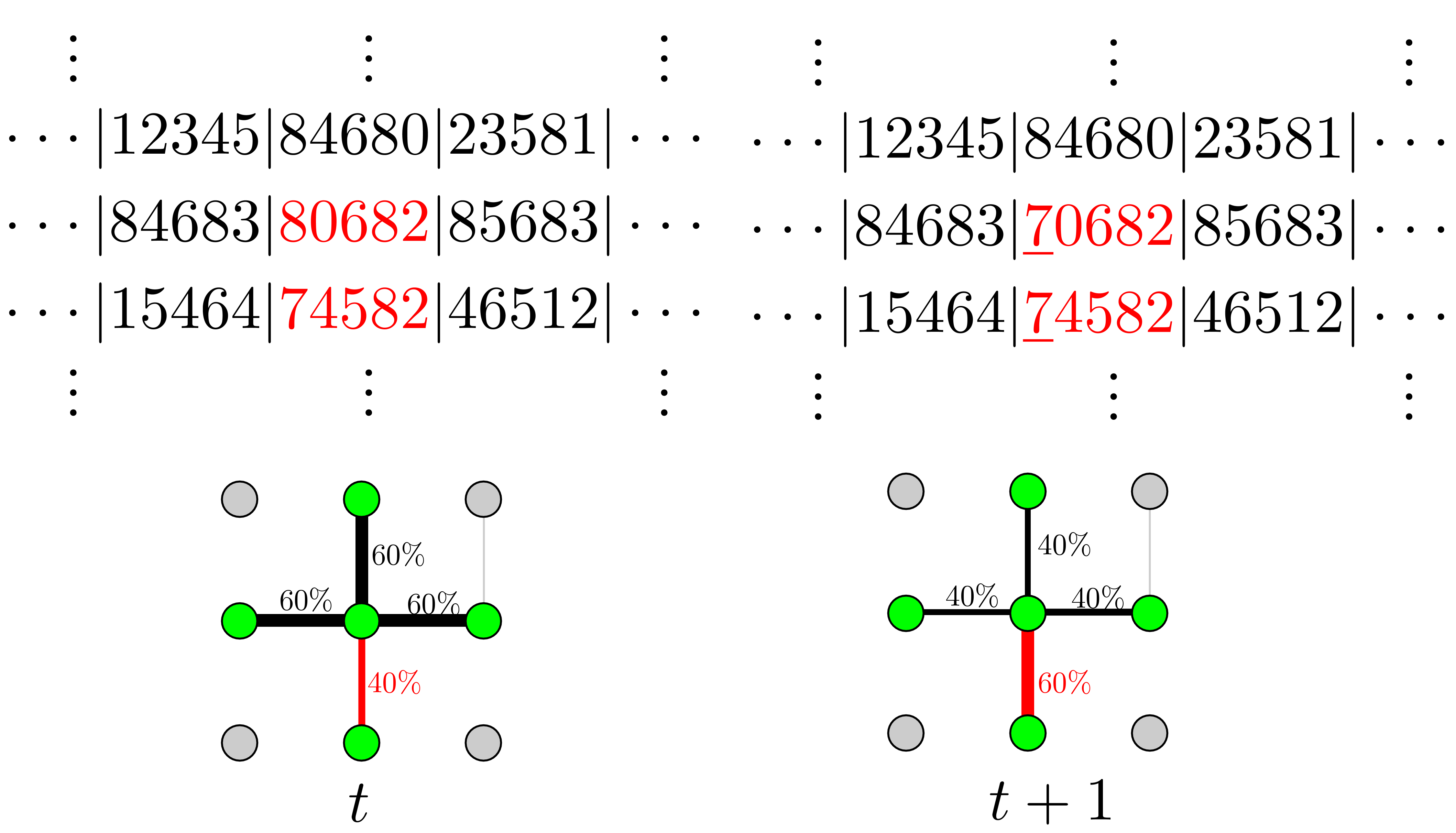} 
\end{figure}
\noindent
Above, the interaction happens at the attribute underlined in the table to the right and give the actual updated state. As one can see, the agent in the middle is the one that is activated. The rest of individual states in the system remain the same so that we find $\left\langle s(t+1)\right\rangle - \left\langle s(t)\right\rangle = -\frac{2}{F}$ as in the illustration above. 

Similarly, for the other extreme case where $\delta_t = \frac{4}{F}$, we have the following pictorial explanation
\begin{figure}[H]
	\centering
	\includegraphics[scale=.7]{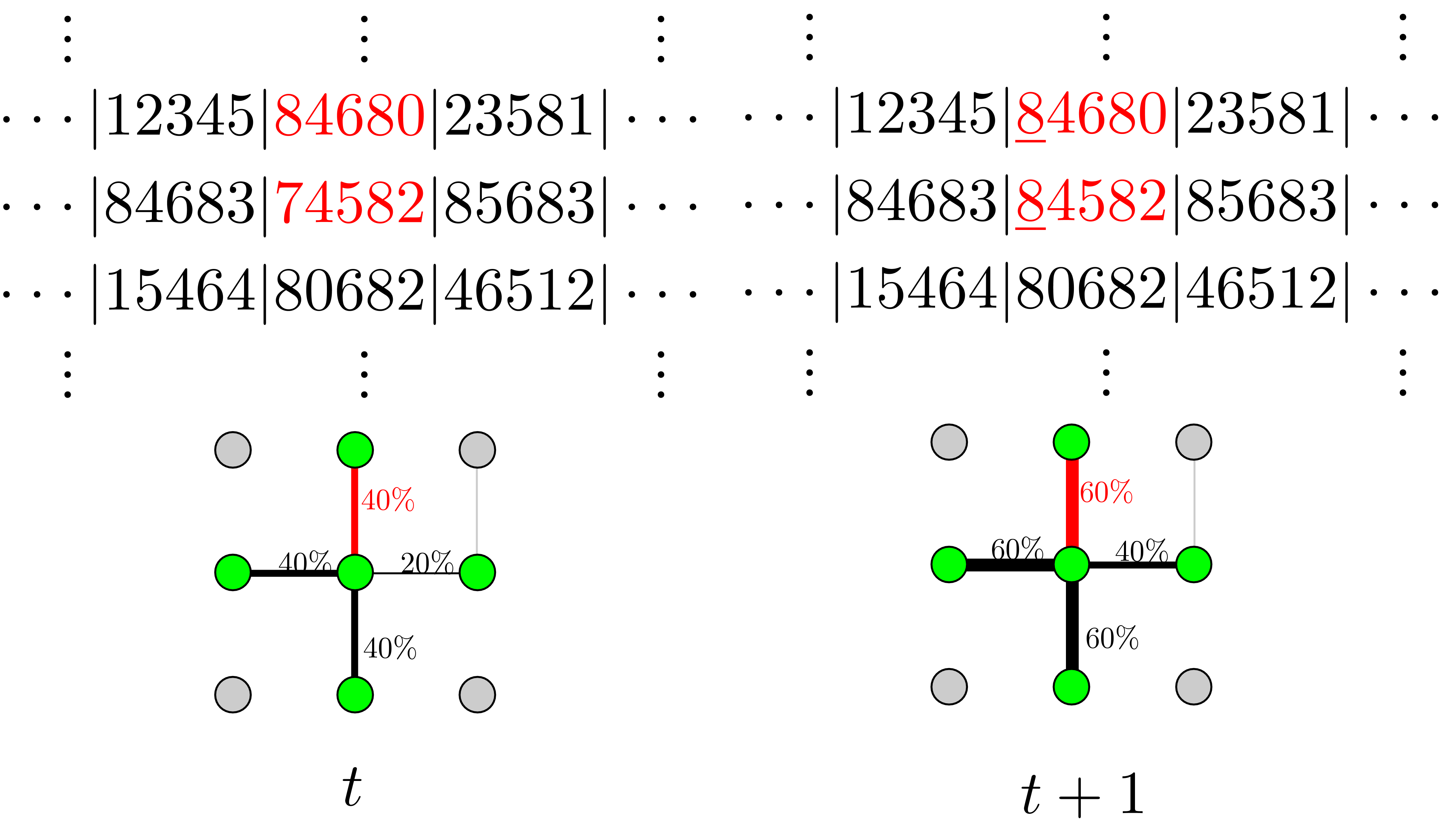} 
\end{figure}
\noindent
The two illustrations above are the extreme ones but the intermediate scenario are expected as well. 

\begin{figure}[H]
	\centering
	\includegraphics[scale=.4]{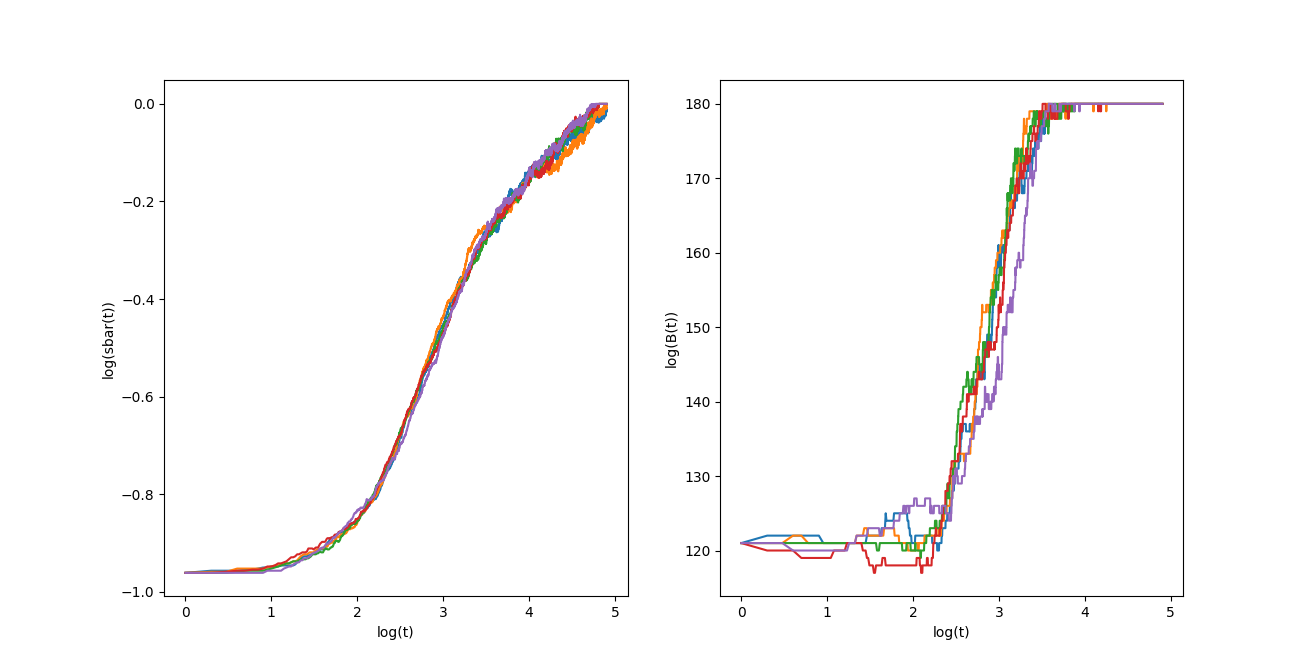}
	\caption{We perform 5 separate simulations from an initial configuration of the system. Each simulation evolves under 80000 times of iterations using our algorithm \ref{Alg1}. To the left, we plot $\log(\left\langle s(t)\right\rangle)$ versus $\log(t)$. To the right we plot $\log(B(t))$ versus $\log(t)$. The real computer time to perform these five simulations is around 45 minutes. We implement this algorithm in python3 on a laptop with a processor: Intel${}^{\textregistered}$ Core${}^{\texttrademark}$ i5 CPU M 560 @ 2.67GHz $\times$ 4.}
	\label{fig:figure1-1}
\end{figure}

By making use of the previous analysis, the generic equation of motion of the system is 
\begin{equation}\label{sbargeneric}
	\left\langle s(t)\right\rangle = \mathsmaller{\frac{1}{N}} + \sum_{n\geq 1}^{t} \delta_n.
\end{equation}
At each time, $\delta_n$ follows a stochastic process and take value in $\{-\frac{2}{F},-\frac{1}{F},0,\frac{1}{F},\frac{2}{F},\frac{3}{F},\frac{4}{F}\}$.
To exhibit the two apparent continuous phase transitions in \eqref{sbargeneric}, concretely we consider a system with the parameters $F=10, N=10$ and $ L = 10$. We run simulations\footnote{See https:$//$sites.google.com/aims.ac.za/nirinaprofessionalwebpage/code-libraries for the python source codes.} 5 times and plot $\left\langle s(t)\right\rangle$ versus time on a log-log scale. We also plot the number of bounds $B(t)$ versus time using the log-log scale. These are shown in figure \ref{fig:figure1-1}. The number of bounds at each time is equivalent to the number of non-zero local degree similarities, $s_{IJ}(t)$ in the system. To explain the different evolution phases of the system, return to \eqref{sbargeneric}. Generically, the number of bounds $B(t)$ during the period $0\leq t\leq 2\Bmax$ remains almost constant. This can be explained from the probability distribution \eqref{eqproba}. The interactions during this period happen only to increase the existing non-trivial local degree of similarities. Among these interactions there is no much expectation of creating new bounds. After this period, in particular for the simulations in figure \ref{fig:figure1-1}, during $2\Bmax \leq t\leq 20 \Bmax$ the system evolves to reach it saturation in terms of $B(t)$. This regime corresponds to an apparent power law of $\left\langle s(t)\right\rangle$ with respect to $t$.  The saturation of $B(t)$ is reached long before the the system reaches its equilibrium. This is the last phase $t\geq 20\Bmax$, it corresponds to the oscillations of $\left\langle s(t)\right\rangle$ as showed in figure \ref{fig:figure1-1}. During these oscillations the extreme values taken by $\delta_n$ are dominant.  
\noindent
This completes our study of the Axelrod model in terms of its quantitative dynamical properties. From now on, we shift our attention to understand its space of culture configurations. In terms of the parameters of the model we recall that the total number of culture configurations in this model is 
\begin{equation}\label{eq:Omega}
\Omega = N^{F\times L^2}.
\end{equation}
However, out of this number, there are only $N^F $configurations that correspond to the monoculture states of the system. That is because for one individual on the lattice, its individual states count in total $N^F$. The rest of the configurations in \eqref{eq:Omega} correspond to multiculture states of the system. To achieve our goal, in the next section, we use techniques from representation theory. We strongly advise the readers unfamiliar with the technicalities and terms in this sections to read the following literatures \cite{curtis90, dmk11,humphreys12} and some of their lists of references.
\section{Culture configurations space and the symmetric group $S_N$}\label{SecCCSN}
For simplicity, we only consider systems at equilibrium where $\left\langle s(t_{\text{eq}})\right\rangle = 1$. By definition this corresponds to systems with monoculture configurations. To achieve this, it is enough to study the space of individual states. We now pursue to our earlier statement that this space is isomorphic to the space $V_N^{\otimes F}$. To make this concrete, we first consider the canonical basis vectors of $V_N^{\otimes F}$. The elements of this basis are given by the set of all possible tensor products of $F$ elements in $\mathcal{B}_N$. We recall the definition of $\mathcal{B}_N$ in equation \eqref{eq:BN}. A bijection between the $\ket{\bq} $'s and the elements of the canonical basis of $V_N^{\otimes F}$ is 
\begin{equation}\label{defIstate2}
\ket{\bq} \equiv \ket{\be_{q_1}\otimes\be_{q_2}\otimes\cdots \otimes\be_{q_F}} .
\end{equation}
Having this definition in mind, from now on we can always think of a state $\ket{\bq} $ as an element in $V_N^{\otimes F}$. Now we are ready to explain how the symmetric group $S_N$ acts on the system. We explain that this action classify the space of culture configurations in a special way. 
\subsection{Action of $S_N$ on the $\ket{\bq} $'s as elements of $V^{\otimes F}_N$}
An elementary fact from representation theory is that there is a natural reducible representation of $S_N$ over the space $V_N$. To see this, consider a permutation $\sigma\in S_N$, the matrix $\Gamma_{V_N}(\sigma)$ representing this element is constructed from the action 
\begin{equation}\label{eq:GVN}
	\Gamma_{V_N}(\sigma) \ket{\be_i} = \ket{\be_{\sigma(i)}} .
\end{equation}
In this way, for all $\sigma\in S_N$ the $\Gamma_{V_N}(\sigma)$'s are $N\times N$ dimensional matrices. 

Next, we want to generalize this natural representation over the space $V_N^{\otimes F}$. The natural generalization of \eqref{eq:GVN} is 
\begin{equation}\label{eq:VNtF}
	\Gamma_{V_N^{\otimes F}}(\sigma)\ket{\be_{q_1}\otimes\be_{q_2}\otimes\cdots \otimes\be_{q_F}} =  \ket{\be_{\sigma(q_1)}\otimes\be_{\sigma(q_2)}\otimes\cdots \otimes\be_{\sigma(q_F)}}
\end{equation}
Clearly, for any $\sigma\in S_N$, equation \eqref{eq:VNtF} is a well defined statement.
Similar to \eqref{eq:GVN} the $\Gamma_{V_N^{\otimes F}}(\sigma)$'s are matrices of dimension $N^F\times N^F$. In fact, it is not difficult to show that the matrices $\Gamma_{V_N}$ and $\Gamma_{V^{\otimes F}_N}$ are related as follow
\begin{equation}
\forall \sigma\in S_N,\quad \Gamma_{V_N^{\otimes F}}(\sigma) = \overset{F}{\overbrace{\Gamma_{V_N}(\sigma)\otimes \Gamma_{V_N}(\sigma)\cdots\otimes \Gamma_{V_N}(\sigma)}}.
\end{equation} 
This equation is in agreement with the fact that the $\Gamma_{V_N^{\otimes F}}(\sigma)$'s must have the dimension $N^F\times N^F$.
In the language of representation theory, actually \eqref{eq:VNtF} defines a representation of the group $S_N$ over the space $V_N^{\otimes F}$. This representation is reducible. In this way, one is interested to find the decomposition of the space $V_N^{\otimes F}$ into invariant subspaces that carry inequivalent irreps of $S_N$. In other words, we are interested to partition the space of individual states into disjoint classes. Only states within the same class can be mixed between them under the group action defined in \eqref{eq:VNtF}. Achieving this classification identifies the different inequivalent classes of monoculture states of the system out of the $N^F$ possibilities.

For a concrete connection to the decomposition problem and our analysis of culture configurations, return to our old notation $\ket{\bq} =  \ket{q_1q_2\cdots q_F}  \in V_N^{\otimes F}$.
In this way, equation \eqref{eq:VNtF} becomes\footnote{Here we just drop the subscript $V_N^{\otimes F}$ on the matrix $\Gamma$ but it should be clear that we are operating in this space.}
\begin{equation}\label{Snaction}
\Gamma(\sigma) \ket{q_1q_2\cdots q_{F-1}q_F} = \ket{\tilde{q}_1\tilde{q}_2\cdots \tilde{q}_{F-1}\tilde{q}_F} ,   \quad \text{with}\quad \tilde{q}_k \equiv \sigma(q_k).
\end{equation}
As a concrete example, consider $F=5,\, N = 10$ and $\sigma = (134)(58)$ with an individual state ${\ket{\bq} } = \ket{52178} $, we find
\begin{equation*}
\Gamma((134)(58))\ket{52178} = \ket{82375} .
\end{equation*}
The definitions \eqref{eq:VNtF} and \eqref{Snaction} are equivalent.
The action of a permutation $\sigma\in S_N$ on the system is to extend one of these two definitions\footnote{This is indeed depending on the notations of the individual states that one chooses to use.} to the whole $L^2$ individual states. It follows that a $\sigma\in S_N$ acts on the system as to relabel simultaneously all the $L^2$ states in the system without changing its gridgraph configuration as in figure \ref{typicalGraph}. In this way it is not difficult to argue that the local degree of similarities $s_{IJ}$ and the average similarities $\left\langle s(t)\right\rangle$ are not affected by this relabelling. This group action can be implemented numerically to verify that indeed $\left\langle s(t) \right\rangle$ is invariant for different culure configurations related by any permutation in $S_N$. We recall the definition of $\left\langle s(t)\right\rangle$ in \eqref{eqsbar}. 

Keeping track of our interest in decomposing the space $V_N^{\otimes F}$ into direct sums of invariant subspaces under this group action, we proceed as follow. We assume that $N\geq F$. Consider a set of $F$ variables $\{x_i\}_{i=1,\cdots,F}$, where $\forall \,k,\,  x_{k}\in \{1,2,3,\cdots, N\}$ satisfying the constraints
\begin{equation*}
x_i \neq x_j\quad \text{if $i\neq j$}. 
\end{equation*}
Consider a diagram of a single row and $F$ adjacent boxes 
\[\overset{F}{\overbrace{\yng(3)\cdots\yng(2)}}.\]
Consider $1\leq n(x_i) \leq F$, where $n(x_i)$ counts the number of each variable $x_i$ to fill these boxes according to the following rules
\begin{algorithm}[H]
	\caption{: Rules to generate diagrams indexing the decomposition of $V_N^{\otimes F}$. These different diagrams identify the inequivalent different monoculture states of the system. }\label{Alg2}
	\begin{algorithmic}[1]
		\Statex \textbullet~\textbf{Parameters:} $\{x_i\},\, n(x_i),\,\text{where}\, \sum_i  n(x_i) =  F$.
		\State Step-1: Start with $x_1$ to fill the $F$ empty boxes with the $n(x_1) = F$ copies of this variable.
		\State Step-2: Use only two variables $x_1$ and $x_2$ to fill the boxes in different possible ways while keeping $1\leq n(x_2)\leq n(x_1)$ and $n(x_1) + n(x_2) = F$.
		\State Step-3: Repeat and iterate one at a time step-2 in terms of the number of variables to use in filling the boxes while keeping $1\leq n(x_F)\leq \cdots \leq n(x_2)\leq n(x_1)$ and $\sum_{i}n(x_i) = F$.
	\end{algorithmic}
\end{algorithm}
\noindent
After finishing this algorithm, we need to identify the first box to the left in $\overset{F}{\overbrace{\Yvcentermath1\Yboxdim{8pt}\yng(3)\cdots\yng(2)}}$ to be the attribute $q_1$ and so on until the last box to the right with $q_F$. Next, substitute the variables $x_i$'s with any of the possible integers they take between $1$ to $N$. 

To make these discussions concrete, a less trivial example is to consider $F=4$, $N\geq 4$. We have four variables $x_1,x_2,x_3$ and $x_4$ to use for the filling of the diagram $\Yvcentermath1\Yboxdim{8pt}\yng(4)$. Following the rules listed in the algorithm \ref{Alg2}, the only possible filled box diagrams we find are
\begin{align}\label{15irreps}
	\young(\xo\xo\xo\xo) && \young(\xo\xo\xto\xto) &&\young(\xo\xto\xtr\xo)\nonumber\\
	\young(\xo\xo\xo\xto) && \young(\xo\xto\xo\xto) &&\young(\xto\xo\xo\xtr)\nonumber\\
	\young(\xo\xo\xto\xo) && \young(\xo\xto\xto\xo) &&\young(\xto\xtr\xo\xo)\\
	\young(\xo\xto\xo\xo) &&\young(\xo\xo\xto\xtr) &&\young(\xto\xo\xtr\xto)\nonumber\\
	\young(\xto\xo\xo\xo) && \young(\xo\xto\xo\xtr) &&\young(\xo\xto\xtr\xfo)\nonumber
\end{align}
These 15 diagrams in \eqref{15irreps} index the inequivalent class of subspaces composing the decomposition of $V^{\otimes 4}_N$. To explain this we need to perform the identification of the boxes with the $q_i$'s followed by the substitution of the $x_i$'s as stated above. In doing so, we find
\begin{enumerate}
	\item For the first diagram that uses only 4 $x_1$'s there are only $N$ possible individual states,
	\item For those diagrams with two types of variables $x_1$ and $x_2$, there are exactly $N(N-1)$ possible states,
	\item For those diagrams with three types of variables $x_1$, $x_2$ and $x_3$, there are exactly $N(N-1)(N-2)$ possible states
	\item At last, with the four types of variables $x_1,\,x_2,\,x_3$ and $x_4$ there are exactly $N(N-1)(N-2)(N-3)$ possible states.
\end{enumerate}

Now, act with the group $S_N$ according to the definitions in \eqref{Snaction}. We argue that the states generated from the diagram $\young(\xo\xo\xo\xo)$ will never mix to any of the states generated from the other 14 diagrams. Similarly to the other inequivalent filled box diagrams. However if one adds the total states from these 15 filled diagrams, there are exactly $N^4$ possible states. This is not an accident because, this is exactly the number of states for an individual with $F=4$ attributes where each attribute takes $N$ different traits. This follows from the identification  between the boxes and the $q_i$'s that follows algorithm \ref{Alg2}. This concrete example demonstrates how the space $V_N^{\otimes 4}$ is decomposed into 15 irreducible subspaces labelled by the diagrams in \eqref{15irreps}. The correct statement is 
\begin{equation*}
	V^{\otimes 4}_N = \,\overset{15}{\underset{i=1}{\bigoplus}}\, V_{d_i},
\end{equation*}
where the $d_i$'s refer to each diagrams in \eqref{15irreps}. The above results is translated as follow in the Axelrod system with $N\geq 4$ and any $L\geq 2$. In this system, the total numbers of monoculture states of the system is $N^4$. However, because of the $S_N$ symmetry present in the system, only these 15 diagrams in \eqref{15irreps} correspond to the inequivalent classes of monoculture states.  
\subsection{Some combinatorics facts vs Algorithm 2:}
In this section, we want to prove the above statement for the cases $N\geq F$ in general. To achieve this, we argue that the problem of counting different filled box diagram configurations of each step in our algorithm \ref{Alg2} can be turned into a well defined combinatorics problem. Following algorithm \ref{Alg2}, assume we are at a step where one has to use the first $k$ variables $\{x_r\}_{r=1,\cdots ,k}$, where $k\leq F$. Recall the constraints $\sum_i n(x_i) = F$ and $1\leq n(x_k)\cdots \leq n(x_2)\leq n(x_1)$. At this stage, filling the $F$ empty boxes with the set of $x_i$'s is exactly equivalent to construct different ways of partitioning a set of $F$  elements into $k$ nonempty sets. The proof of this counting equivalence between the different diagrams and partition of a set is not difficult to do. So we will leave it as an exercise to interested readers. Our main interest here is to exploit it and prove the general statement about the decomposition of $V_N^{\otimes F}$. Following \cite{weissteinSNSK} the Stirling number of second kind denoted by ${F\brace k} $ gives exactly the number of ways of partitioning a set of $F$ elements into $k$ nonempty sets. To complete all the steps in our algorithm \ref{Alg2}, we let $k=1,\cdots,F$. In this way, we are guaranteed that all the possible inequivalent filled box diagrams are achieved. Accordingly, the total number of inequivalent diagrams is 
\begin{equation}
	B_F = \sum_{k=1}^F {F \brace k}.\label{eq:sumofFbracek}
\end{equation}
This sum is a among the well known identities that the ${F\brace k}$'s satisfy in combinatorics. $B_F$ is the so-called Bell number. Another identity we need to finalize our discussion of decomposing the space $V_N^{\otimes F}$ is 
\begin{equation}
	\sum_{k=1}^F {F\brace k} (x)_k = x^F. \label{eq:sumtoxtoF}
\end{equation} 
In this sum, $x$ is an integer such that $x\geq F$. The symbol $(x)_k$ is the falling factorial 
\begin{equation}
	(x)_k = x(x-1)(x-2)\cdots (x+2-k)(x+1-k).
\end{equation} 
Using \eqref{eq:sumtoxtoF}, by substituting $x$ to $N$, we find exactly the correct dimensions between $V^{\otimes F}_N$ and the components composing its direct sum decomposition. This completes our proof and find that
\begin{equation}\label{decomposisionVFN}
	V^{\otimes F}_N = \,\overset{B_F}{\underset{i=1}{\bigoplus}}\, V_{d_i},
\end{equation}
where the $d_i$'s are indexing the diagrams produced, following our algorithm \ref{Alg2}. 

Instead of using the box diagrams we proposed in this work, there are other diagrams known as Dickau diagrams directly associated to the Bell number $B_F$. The Dickau diagrams associated to $B_F$ can be used alternatively to label the irreps of $S_N$ in the decomposition \eqref{decomposisionVFN}. If one has to reconstruct the Dickau diagrams from our filled box diagrams, the procedure is as follow. Construct a $F-$regular polygon and identify from inside each corners by dots. Start from the top corner and its dot, we associate this to the last box to the right in our diagram. Continue this association similarly to the other corners of the polygon and their respective dots. Perform this by going clockwise such that we go from right to the left using the remaining boxes. Then, if any boxes contain the same variable $x_i$, in the Dickau diagrams link their respective associated dots with a segment. To illustrate this, return to our earlier example and find
\begin{figure}[H]
	\centering
	\includegraphics[scale=0.83]{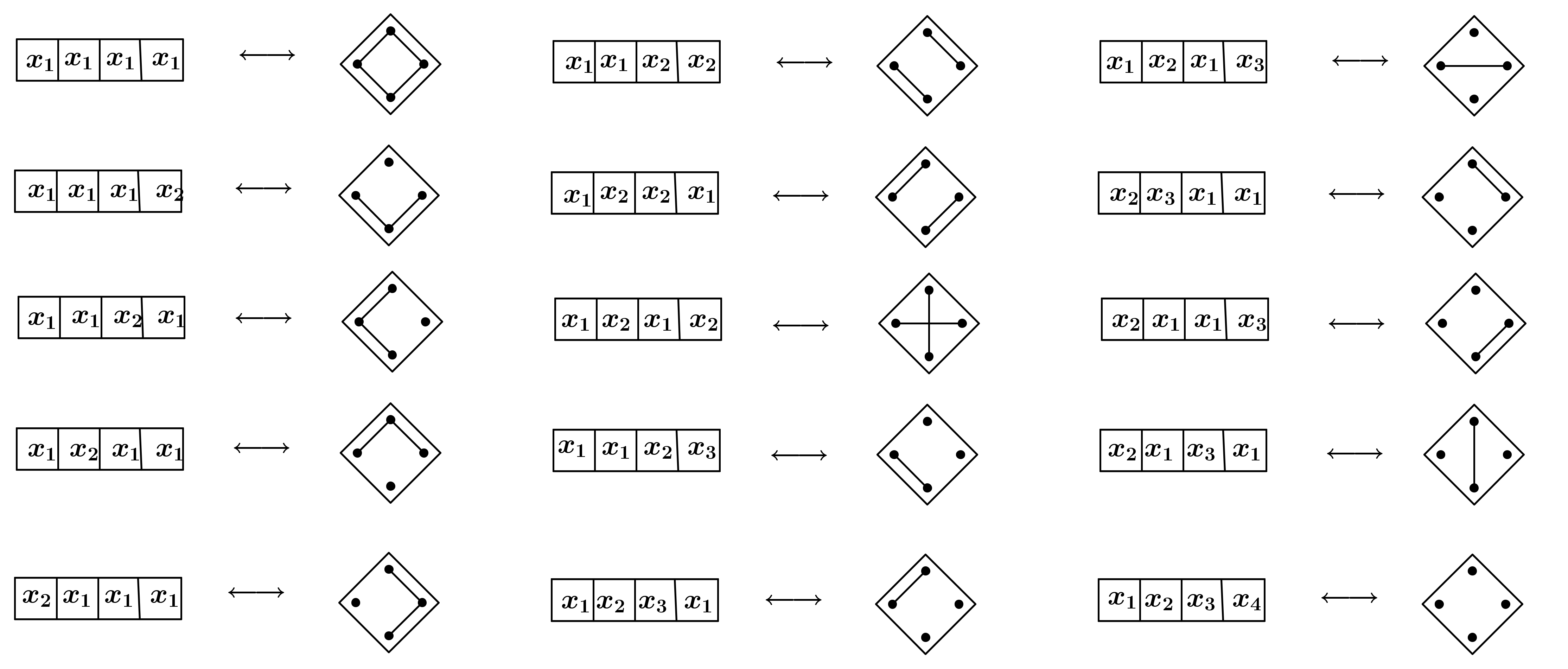}
	\caption{Example illustrating the relationship between the Dickau diagrams and our filled box diagrams. Here $F=4$ so that the Bell number $B_4=15$.}
	\label{fig:belldia}
\end{figure}

This complete our discussion for system with monoculture states, $\left\langle s(t_{\text{eq}})\right\rangle =1$. The Dickau diagrams as illustrated in figure \ref{fig:belldia} can be used to index each class of of different inequivalent monoculture states of the Axelrod system. The case for different equilibrium culture configurations of the system where $\left\langle s(t_{\text{eq}})\right\rangle<1$ can be treated similarly to our analysis above. These are called multiculture states of the system. We claim that the number of different classes of inequivalent culture configurations must involve a power of the Bell number $B_F$.
\section{Conclusions}\label{concl}
To conclude our discussion, one checks that our results can be used to prove the hypothesis and results observed in the original paper by Axelrod \cite{Axelrod97}. In fact, we consider that the results we reported here are more general in the following sense. We derive our results analytically so that predictions based on the model beyond the existing numerical results are provided. Our finding also are in agreement with the existing literatures - inspired by the Axelrod model -  we cited in this work. In addition to these, we have established the analysis of the space of culture configurations in this model. As far as we know this has never been considered in any literature. This analysis of the space of culture configurations exploits the existence of $S_N$ symmetry and leads us to the following discovery. We establish a connection between group representation theory and certain combinatorics problem to the Axelrod model. This is summarized in the decomposition of - see \eqref{decomposisionVFN} - the space $V^{\otimes F}_N$ identical to the space of individual's culture. Even though we did it for the monoculture case, $\left\langle s(t)\right\rangle = 1$, we argue that one can extend our approach to all different multiculture cases. 

There are few directions that can be considered for future extensions of this work. First, one can start to include external authorities and study their impacts on the model. A motivation for this study is to pursue the idea of social science/statistical physics correspondence. Our proposal for this problem is to make an analogy with an external magnetic field on a spin lattice - just like the Ising model. This entails that certain Hamiltonian formulation of the system must be provided. However, a simple alternative extension to our study is to change boundary conditions and study their effects. A question we like to answer in the model without any external authorities is to characterize the system when reaching multicultural equilibrium. Long range interactions, and multiple activations within the dynamics of the system are also interesting points to study. In addition, the study of the generic motion and the appearance of a phase that follows a power law is needed to be fully understood. We argue there should be certain analytic way of explaining these behaviours. These are among few topics we propose for future study.

\section*{Acknowledgement}
We want to express our gratitude to the AIMS family members to make this work possible. We also thank Robert de Mello Koch, Konstantinos Zoubos, Cynthia Ramiharimanana, Georg Anegg and Alberto Cazzaniga for useful discussions and comments. Yabebal Tadesse Fantaye is supported by AIMS-ARETE chair. Nirina M. Hasina Tahiridimbisoa is supported by AIMS.
\appendix
\section{Some proofs and numerical support of the typical initial configurations}\label{AppendixA}
To prove \eqref{sbart0} we consider the probability \eqref{eqproba}. This probability gives an initial distribution of similarities in the system. Let $X_{n}\big|_{n=0\cdots,F}$ be the random variables counting the proportion of similarity $s = \frac{n}{F}$ in the system. By definition, we find
\begin{equation}
X_{n} = \Bmax\cdot P(s= \mathsmaller{\frac{n}{F}}).
\end{equation}
Accordingly, computing $\left\langle s(t)\right\rangle_{t=0}$ using \eqref{eqsbar} becomes simply the evaluation of a mean value of the Binomial distribution and find  
\begin{align}
\left\langle s(t)\right\rangle_{t=0} &=\frac{1}{\Bmax}\sum_{n = 0}^{F} \frac{n}{F} \Bmax \cdot P(\frac{n}{F}),\nonumber\\
&= \frac{1}{N}.
\end{align}
Since each $X_{n}\big|_{n=0,\cdots, F}$ are independent, one computes independently their respective variances to find 
\begin{equation}\label{Stdiniconfig}
\varDelta X_n^2 = \Bmax  P(\mathsmaller{\frac{n}{F}})(1-P(\mathsmaller{\frac{n}{F}})).
\end{equation}
In return, these variances in \eqref{Stdiniconfig} can be used to estimate the standard deviation of $\left\langle s(t)\right\rangle_{t=0}$ and find 
\begin{equation}\label{varestimated}
\Sigma_{\left\langle s(t)\right\rangle_{t=0}} = \frac{\sqrt{\sum_{n=1}^{F} \varDelta X_n^2 - \varDelta X_0^2}}{\Bmax}.
\end{equation}
The numerical evidence supporting the above discussions is plotted in the following figure.
\begin{figure}[H]
	\centering
	\label{fig:histiniconfig}
	\includegraphics[scale=.5]{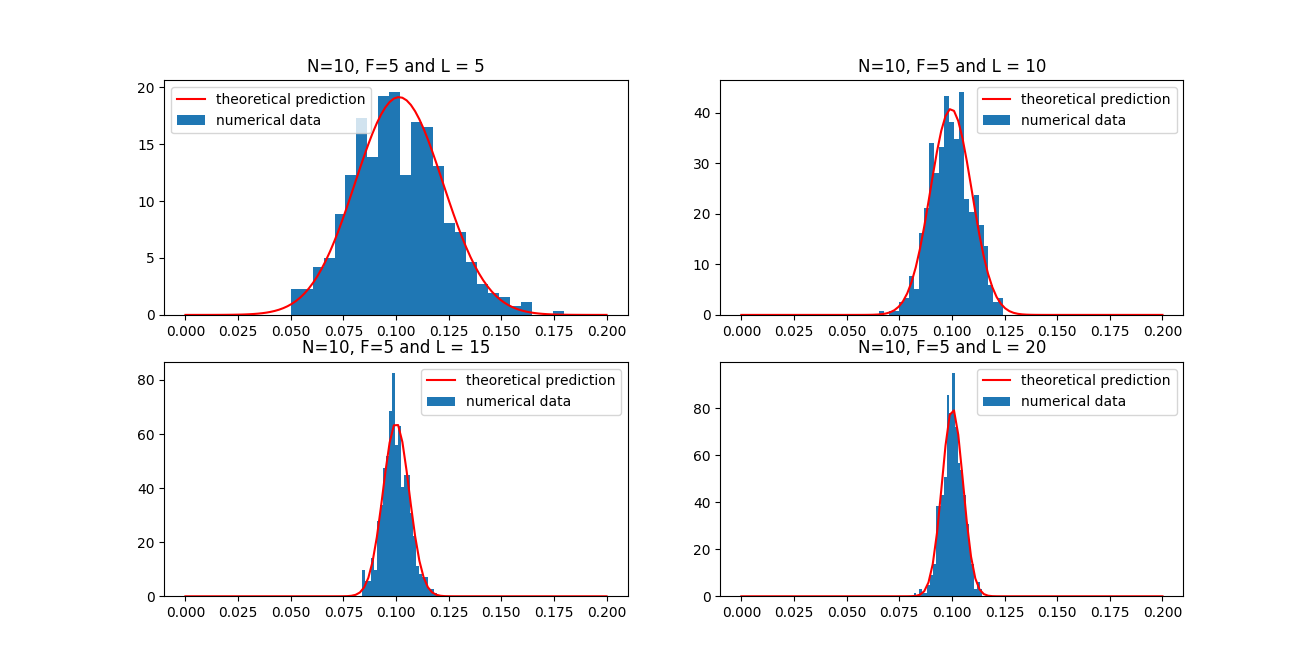}
		\caption{Distribution of a typical configuration with $\left\langle s(t)\right\rangle_{t=0}=\frac{1}{N}$ and standard deviation given as in equation \eqref{varestimated}. These data were generated from 1000 initialization samples of the system with the parameters $N=10, F=5$ and respectively $L=[5,10,15,20]$ as shown in the subplots above.}
\end{figure}

%%-----------------------------------------------------------------------------------------------------------%
%\renewcommand{\bibname}{References}
%\nocite{*}
\bibliographystyle{hunsrt}
\bibliography{Axelrodpaperbib}

\begin{thebibliography}{10}

\bibitem{homans61}
George~C. Homans.
\newblock {\em Social behavior: Its elementary forms}.
\newblock New York, Harcourt, Brace \& World, 1961.

\bibitem{triandis94}
H.C. Triandis.
\newblock {\em Culture and Social Behavior}.
\newblock McGraw-Hill series in social psychology. McGraw-Hill, 1994.

\bibitem{gilbert95}
Nigel Gilbert and Rosaria Conte.
\newblock {\em Artificial Societies: The Computer Simulation of Social Life}.
\newblock Taylor \& Francis, Inc., Bristol, PA, USA, 1995.

\bibitem{epstein96}
Joshua~M. Epstein and Robert Axtell.
\newblock {\em Growing Artificial Societies: Social Science from the Bottom
  Up}.
\newblock The Brookings Institution, Washington, DC, USA, 1996.

\bibitem{Axelrod97}
Robert Axelrod.
\newblock The dissemination of culture: A model with local convergence and
  global polarization.
\newblock {\em Journal of Conflict Resolution}, 41(2):203--226, 1997,
  https://doi.org/10.1177/0022002797041002001.

\bibitem{Jin01}
Emily~M. Jin, Michelle Girvan, and M.~E.~J. Newman.
\newblock Structure of growing social networks.
\newblock {\em Phys. Rev. E}, 64:046132, Sep 2001.

\bibitem{miller01}
Miller McPherson, Lynn Smith-Lovin, and James~M Cook.
\newblock Birds of a feather: Homophily in social networks.
\newblock {\em Annual Review of Sociology}, 27(1):415--444, 2001,
  https://doi.org/10.1146/annurev.soc.27.1.415.

\bibitem{newman02}
M.~E.~J. Newman.
\newblock Assortative mixing in networks.
\newblock {\em Physical Review Letters}, 89(20):208701, October 2002.

\bibitem{grimm05}
V.~Grimm and S.F. Railsback.
\newblock {\em Individual-based Modeling and Ecology}.
\newblock EBSCO ebook academic collection. Princeton University Press, 2005.

\bibitem{goldstone05}
M.~A.~Janssen Robert L.~Goldstone.
\newblock Computational models of collective behavior.
\newblock {\em Trends in cognitive sciences}, 9:424--30, 2005.

\bibitem{jasmine06}
Andrew~Tomkins Ravi~Kumar, Jasmine~Novak.
\newblock Structure and evolution of online social networks.
\newblock {\em KDD '06 Proceedings of the 12th ACM SIGKDD international
  conference on Knowledge discovery and data mining}, pages 611--617.

\bibitem{vasques07}
F.~Vazquez and S.~Redner.
\newblock Non-monotonicity and divergent time scale in axelrod model dynamics.
\newblock {\em EPL (Europhysics Letters)}, 78(1), April 2007.

\bibitem{goldstone08}
Robert~L. Goldstone, Michael~E. Roberts, and Todd~M. Gureckis.
\newblock Emergent processes in group behavior.
\newblock {\em Current Directions in Psychological Science}, 17(1):10--15,
  2008, https://doi.org/10.1111/j.1467-8721.2008.00539.x.

\bibitem{davidletal09}
David~Lazer et~al.
\newblock Life in the network: the coming age of computational social science.
\newblock {\em HHS Author Manuscripts}, 323(5915):721--723, Feb 06 2009.

\bibitem{buchanan}
Mark Buchanan.
\newblock {\em The Social Atom: Why the Rich Get Richer, Cheaters Get Caught,
  and Your Neighbor Usually Looks Like You}.
\newblock Bloomsbury USA, 2007.

\bibitem{steven99}
Steven~N. Durlauf.
\newblock How can statistical mechanics contribute to social science?
\newblock {\em Proc Natl Acad Sci U S A}, 96(19):10582 -- 10584, Sep 14 1999.

\bibitem{CCFVL09}
Claudio Castellano, Santo Fortunato, and Vittorio Loreto.
\newblock Statistical physics of social dynamics.
\newblock {\em Rev. Mod. Phys.}, 81:591--646, May 2009.

\bibitem{gandica13}
Y.~Gandica, E.~Medina, and I.~Bonalde.
\newblock A thermodynamic counterpart of the axelrod model of social influence:
  The one-dimensional case.
\newblock {\em Physica A: Statistical Mechanics and its Applications},
  392(24):6561 -- 6570, 2013.

\bibitem{genzor15}
Jozef Genzor, Vladimír Bužek, and Andrej Gendiar.
\newblock Thermodynamic model of social influence on two-dimensional square
  lattice: Case for two features.
\newblock {\em Physica A: Statistical Mechanics and its Applications}, 420:200
  -- 211, 2015.

\bibitem{DavidTong12}
David Tong.
\newblock Lectures on statistical physics.
\newblock page 190, 2011-2012,
  http://www.damtp.cam.ac.uk/user/tong/statphys.html.

\bibitem{vfrjones93}
V.F.R. Jones.
\newblock The potts model and the symmetric group.
\newblock {\em Subfactor: Proceedings of the Taniguchi Symposium on Operator
  Algebras (Kyuzeso, 1993), World Sci. Publishing, River Edge, NH}, pages
  259--267, 1994.

\bibitem{halverson05}
Tom Halverson and Arun Ram.
\newblock Partition algebras.
\newblock {\em European Journal of Combinatorics}, 26(6):869 -- 921, 2005.
\newblock Combinatorics and Representation Theory.

\bibitem{zajj14}
Zajj Daugherty and Rosa Orellana.
\newblock The quasi-partition algebra.
\newblock {\em Journal of Algebra}, 404:124 -- 151, 2014.

\bibitem{lanchier12}
Nicolas Lanchier.
\newblock The axelrod model for the dissemination of culture revisited.
\newblock {\em Ann. Appl. Probab.}, 22(2):860--880, 04 2012.

\bibitem{maldacena97}
Juan~Martin Maldacena.
\newblock {The Large N limit of superconformal field theories and
  supergravity}.
\newblock {\em Int. J. Theor. Phys.}, 38:1113--1133, 1999, hep-th/9711200.
\newblock [Adv. Theor. Math. Phys.2,231(1998)].

\bibitem{corley01}
Steve Corley, Antal Jevicki, and Sanjaye Ramgoolam.
\newblock {Exact correlators of giant gravitons from dual N=4 SYM theory}.
\newblock {\em Adv. Theor. Math. Phys.}, 5:809--839, 2002, hep-th/0111222.

\bibitem{dmk11}
Robert de~Mello~Koch, Matthias Dessein, Dimitrios Giataganas, and Christopher
  Mathwin.
\newblock {Giant Graviton Oscillators}.
\newblock {\em JHEP}, 10:009, 2011, 1108.2761.

\bibitem{dMKH16}
Robert de~Mello~Koch, Nirina~Hasina Tahiridimbisoa, and Christopher Mathwin.
\newblock {Anomalous Dimensions of Heavy Operators from Magnon Energies}.
\newblock {\em JHEP}, 03:156, 2016, 1506.05224.

\bibitem{NMHT}
Abdelhamid Mohamed Adam~Ali, Robert de~Mello~Koch, Nirina~Hasina
  Tahiridimbisoa, and Augustine Larweh~Mahu.
\newblock {Interacting Double Coset Magnons}.
\newblock {\em Phys. Rev.}, D93(6):065057, 2016, 1512.05019.

\bibitem{dmk17}
Robert de~Mello~Koch, Phumudzo Rabambi, Randle Rabe, and Sanjaye Ramgoolam.
\newblock {Counting and construction of holomorphic primary fields in free CFT4
  from rings of functions on Calabi-Yau orbifolds}.
\newblock {\em JHEP}, 08:077, 2017, 1705.06702.

\bibitem{dmk18}
Robert de~Mello~Koch and Lwazi Nkumane.
\newblock {From Gauss Graphs to Giants}.
\newblock {\em JHEP}, 02:005, 2018, 1710.09063.

\bibitem{castellano00}
Claudio Castellano, Matteo Marsili, and Alessandro Vespignani.
\newblock Nonequilibrium phase transition in a model for social influence.
\newblock {\em Phys. Rev. Lett.}, 85:3536--3539, Oct 2000.

\bibitem{rodriguez10}
Arezky~H. Rodr\'{\i}guez and Y.~Moreno.
\newblock Effects of mass media action on the axelrod model with social
  influence.
\newblock {\em Phys. Rev. E}, 82:016111, Jul 2010.

\bibitem{curtis90}
C.W. Curtis and I.~Reiner.
\newblock {\em Methods of Representation Theory: With Applications to Finite
  Groups and Orders}.
\newblock Number v. 1 in Wiley Classics Library. Wiley, 1990.

\bibitem{humphreys12}
J.E. Humphreys.
\newblock {\em Introduction to Lie Algebras and Representation Theory}.
\newblock Graduate Texts in Mathematics. Springer New York, 2012.

\bibitem{weissteinSNSK}
Eric~W. Weisstein.
\newblock Stirling number of the second kind.
\newblock Visited on 28/02/18.

\end{thebibliography}
%%%%------------------------------------------------------------------------
\end{document}